\documentclass[sigconf]{acmart}

\usepackage{booktabs} 
\usepackage{tikz}
\usepackage{csquotes}
\usetikzlibrary{decorations.pathreplacing, angles, quotes, matrix, positioning,
  patterns}
\usepackage[linesnumbered]{algorithm2e}

\setcopyright{rightsretained}

\acmConference[ELS'19]{the 12th European Lisp Symposium}{April 01--02 2019}{%
  Genova, Italy}
\acmISBN{978-2-9557474-3-8}
\acmDOI{}
\setcopyright{rightsretained}
\copyrightyear{2019}

\acmSubmissionID{123-A12-B3}

\begin{document}
\title{Working with first-order proofs and provers}
\subtitle{}

\author{Michael Raskin \and Christoph Welzel}
\authornote{
The project has received funding from the European Research Council (ERC) under the European Union’s Horizon 2020 research and innovation programme under grant agreement No 787367
}
\orcid{0000-0002-6660-5673, XXXX-XXXX-XXXX-XXXX}
\affiliation{%
  \institution{Technical University of Munich}
  \streetaddress{Boltzmannstraße, 3}
  \city{Garching bei M\"unchen}
  \postcode{85748}
}
\email{raskin@mccme.ru, welzel@in.tum.de}

\renewcommand{\shortauthors}{M. Raskin, C. Welzel}

\begin{abstract}
Verifying software correctness has always been an important and complicated
task. Recently, formal proofs of critical properties of algorithms and even
implementations are becoming practical.
Currently, the most powerful automated proof search tools use first-order logic
while popular interactive proof assistants use higher-order logic.

We present our work-in-progress set of tools
that aim to eventually provide
a usable first-order logic computer-assisted proof environment.
\end{abstract}

%
%



\maketitle

\section{Introduction}

Complete formal verification of algorithms and their implementations
is becoming more widely applicable.
Probably the most general approach is construction of formal proofs
in a chosen theory.
Interactively constructed formal proofs often use one of the popular
higher order logics, such as Calculus of Coinductive Constructions
in case of Coq\cite{Coq}
or the chosen higher-order logic of Isabelle/HOL\cite{Isabelle}.
At the same time, a lot of progress in automated reasoning
is achieved in the field of first-order logic.
For example, the Conference on Automated Deduction (CADE)
has an automated theorem prover (ATP) competition,
called CADE ATP System Competition (CASC)\cite{CASC}.
Satallax\cite{Satallax}, the leading higher-order ATP
according to the CASC results\cite{CASC-results},
uses E prover\cite{E} (one of the leading first-order provers)
for some tasks.
On the other hand,
CoqHammer\cite{CoqHammer},
a tool that aims to partially automate 
interactive construction of proofs with Coq,
also uses translation into first-order logic and multiple first-order 
automated provers.

We want to be able to verify statements about distributed algorithms where
direct application of generic ATP systems
might still be impractical.
To that aim, we create first-order specifications, and
use domain knowledge to write or generate
proofs as sequences of lemmas,
while automated theorem provers verify implications.
As CASC competitions have popularized the unified input format
of the Thousands of Problems for Theorem Provers
(TPTP) collection\cite{TPTP},
using multiple ATP systems does not require
any changes in the proof format.
To support this kind of exploration, we develop supporting tooling
for managing the specification, preparing the list of lemmas,
and interacting with the proof system.

While it is too early to draw any conclusions from our ongoing
experiments with representing properties of distributed systems
in the first-order logic,
we want to present the supporting tooling used in this research.

\section{Our tooling}

\subsection{Data formats}

All of our tooling uses TPTP for all the output and most of the proof input.
We use the SyntaxBNF file from the TPTP distribution
(Backus–Naur form of the TPTP format definition) and translate it
into Esrap\cite{Esrap} rules to parse the format.
It turns out that unambiguity of the official TPTP BNF specification
allows us to order the parsing rules in a way compatible
with packrat parsing\cite{packrat}.
More specifically, in every alternation rule
the first (after reordering) successful option can be taken.

However, the formal specifications of the systems in question
contain large amounts of similar statements.
These specifications are generated programmatically.
Currently we do not want to make lasting decisions
about the structure of the specifications we will work with,
so the generating code is written in Common Lisp and refactored
according to the current specification in question.

The proof itself contains additional definitions and lemmas,
and various instructions such as advice to prove some lemma
by case analysis (with a list of cases provided).
We use TPTP Process Instructions (TPI) extension\cite{TPI}
of the TPTP syntax to encode the additional 
imperative instructions related to lemma list processing.

\subsection{Global workflow}

First of all, we need to generate the axioms describing our formal
specification.

To validate that the axioms describe the intended model,
we generate a test run by evaluating the transition rules.
We have code that can evaluate a first-order formula
on an incomplete model, if the fixed part of the model
is enough to determine the formula value easily.
The generated runs are validated in two ways: by manual inspection,
and by verifying that an automated theorem prover
given this run and the full specification does not find a contradiction
in reasonable time.

At our current stage of exploration,
the next step involves writing a list of lemmas
(and instructions for their preprocessing)
that should be sufficient to prove the desired condition.

The last step is verifying that a list of lemmas constitutes
a correct proof.
Of course, in practice this step is performed in parallel
with the previous one.
A part of verification is performed,
then the proof is updated
to avoid the problems observed during verification attempt.
The verification attempts are usually started inside the part of the proof
currently of interest, and stop when some lemma cannot be proved.

Even after the last step of proof verification our tooling 
can offer some further support.
We have some utilities for analysing and visualising 
the output of an ATP system.

The system currently does not provide any dedicated user interface.
It can be used either from a Common Lisp REPL, or via wrapper shell scripts
invoking necessary operations.

\subsection{Structure of an example model}

The examples will be related to one possible encoding of the Dijkstra's
mutual exclusion protocol\cite{Dijkstra} executed on a single CPU
with multiple time-sharing processes (as illustrated in Algorithm
\ref{alg:dijkstra}).
\begin{algorithm}
  \SetKw{KwGoTo}{go to}
  \SetKw{KwTo}{to}
  \SetKw{KwContinue}{continue}
  \Begin{
    $Stealable_{i}\longleftarrow false$\;
    \If{$turn\neq i$}{
      $Outside_{i}\longleftarrow true$\;
      \If{$Stealable_{turn} = true$}{
        $turn \longleftarrow i$\;
      }
      \KwGoTo 3\;
    }\Else{
      $Outside_{i}\longleftarrow false$\;
      \For{$counter_{i}\longleftarrow 1$ \KwTo $n$}{
        \lIf{$counter_{i} = i$}{\KwContinue}
        \lIf{$Outside_{counter_{i}} = false$}{\KwGoTo 3}
      }
    }
    $<$critical section$>$\;
    \label{line:cs}
    $Outside_{i} \longleftarrow true$\;
    $Stealable_{i}\longleftarrow true$\;
    $<$remainder of cycle$>$\;
    \KwGoTo 2\;
  }
  \caption{Dijkstra's algorithm for process $i$ with $n$ parallel processes.}
  \label{alg:dijkstra}
\end{algorithm}
In general, this model has a set of agents switching between states, and
local variables. We automatise a linearly ordered discrete time model which uses
an initial value $initial$ and a function $next\_moment(T)$ which
\enquote{advances} the time one step. The state of agent $A$ at moment $T$
is represented by the function value $active\_state(T, A)$.
There are also some other per-agent variables (and a global $turn$
variable), modelled in the same way, e.g. $counter(T, A)$ which
gives the value of the variable $counter$ of agent $A$ at time $T$.
For a single-CPU multi-process execution we can assume that only one agent
at a time can change its state or variables, and denote this agent as
$active\_agent(T)$. We want to avoid a situation where two agents execute
the critical section (i.e. have the active state equal
to $criticalSection$ which represents line \ref{line:cs} in Algorithm
\ref{alg:dijkstra}) at the same time.

The safety-critical part of the Dijkstra's mutual exclusion protocol
consists of an agent declaring its intent to enter the critical section,
and checking that no other agent has also declared the same intention.

To avoid encoding a full theory with induction,
one can start with proving just the inductive step:
define an invariant then prove that this invariant 
at some moment implies the same invariant at the next moment of time,
and that the invariant implies safety.
The reason to delay encoding the full proof by induction is that the
most natural ways to encode induction axiomatically
require an infinite number of axioms.
This is often called ``induction axiom schema'' --- for every formula 
expressing a predicate, there is an axiom.
This axiom claims that proving 
the base case and the induction step
for the property in question
is enough to verify the property for all natural numbers.

\subsection{Representation of the proof}

In the TPTP format each statement is given a role;
we parse the list of statements and look at their roles.
Axioms are introduced in the specification, and can be used directly.

\enquote{Checked definitions} can be introduced in the proof;
they are axioms that introduce simple abbreviations,
extending the theory in a conservative way.
We verify that they define a single name in terms of previously
seen names, and use these definitions as axioms.
For example, we can define the safety condition.
\begin{verbatim}
fof(define_safety_for, checked_definition,
      ![T,A1,A2]: (safe_for(T,A1,A2)<=>(
         (active_state(T,A1)=criticalSection
           & active_state(T,A2)=criticalSection)
         => A1=A2))).
\end{verbatim}
This formula, which can be also rewritten in the usual notation as 
$\forall T,A_1,A_2: (safe\_for(T,A_1,A_2) \Leftrightarrow
  ((state(T,A_1)=state(T,A_2)=criticalSection)\Rightarrow A_1=A_2))$,
says that a moment $T$ is safe for a pair of agents $A_1$ and $A_2$
if either at least one of the agents is out of the critical section
or they actually are the same agent.
This reflects a part of the desired property
that two different agents should not
execute the critical section simultaneously.
The full safety condition is defined by requiring this property
to hold for each pair of agents.

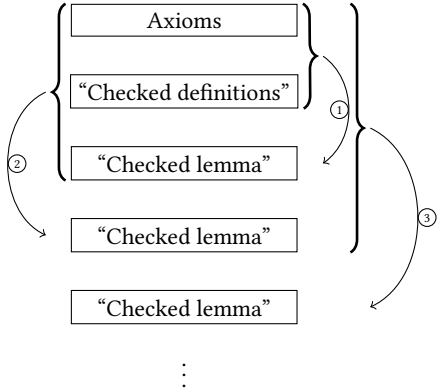
\begin{figure}
  \caption{Structure and proof-steps of our tool.}
  \label{fig:structure-of-proof}
  \begin{tikzpicture}
    \node [draw, rectangle, minimum width = 3cm] (axioms) {Axioms};
    \node [draw, rectangle, minimum width = 3cm, below=0.5 of axioms]
      (checkeddefinitions) {\enquote{Checked definitions}};
    \node [draw, rectangle, minimum width = 3cm, below=0.5 of
      checkeddefinitions] (checkedlemma1) {\enquote{Checked lemma}};
    \node [draw, rectangle, minimum width = 3cm, below=0.5 of checkedlemma1]
      (checkedlemma2) {\enquote{Checked lemma}};
    \node [draw, rectangle, minimum width = 3cm, below=0.5 of checkedlemma2]
      (checkedlemma3) {\enquote{Checked lemma}};

    \draw [ decoration = { brace, amplitude = 5pt, raise = 2pt }, decorate,
    line width = 1pt ] (axioms.north east) --
    node [right, xshift = 6pt] (lemma1) {} (checkeddefinitions.south east);
    \draw [ decoration = { brace, amplitude = 5pt, raise = 2pt, mirror },
    decorate, line width = 1pt ] (axioms.north west) --
    node [left, xshift = -6pt] (lemma2) {} (checkedlemma1.south west);
    \draw [ decoration = { brace, amplitude = 5pt, raise = 20pt }, decorate,
    line width = 1pt ] (axioms.north east) --
    node [right, xshift = 24pt] (lemma3) {} (checkedlemma2.south east);

    \node (checkedlemma1anchor) at (checkedlemma1.east-|lemma1) {};
    \node (checkedlemma2anchor) at (checkedlemma2.west-|lemma2) {};
    \node (checkedlemma3anchor) at (checkedlemma3.west-|lemma3) {};

    \draw [->, bend left = 55] (lemma1.center) to node[left, draw, circle,
    inner sep = 1pt] {\tiny 1} (checkedlemma1anchor.center);
    \draw [->, bend right = 65] (lemma2.center) to node[right, draw, circle,
    inner sep = 1pt] {\tiny 2} (checkedlemma2anchor.center);
    \draw [->, bend left = 65] (lemma3.center) to node[right, draw, circle,
    inner sep = 1pt] {\tiny 3} (checkedlemma3anchor.center);

    \node [below = 0.2 of checkedlemma3] (dots) {\vdots};
  \end{tikzpicture}
\end{figure}

\enquote{Checked lemmas} constitute the main part of the proof.
Every such lemma is first given to an automated prover as
a conjecture to prove, using the axioms and previous lemmas.
If the prover reports a success,
the lemma can be used as an axiom during the following steps.
These steps are
illustrated in Figure \ref{fig:structure-of-proof}.

For example, the following lemma is proved as a part of a case by case analysis.
\begin{verbatim}
fof(safety_conditions_local_cases, checked_lemma,
  ![T,A,B]:( ~passed(T,A,B)
    | (passed(T,A,B) & ~passed(T,B,A))
    | (passed(T,A,B) & passed(T,B,A)))).
\end{verbatim}
This lemma uses the predicate $passed$, that is defined to mean that agent $A$ 
has declared its intent to enter the critical section and has already 
checked that agent $B$ had not declared such at intent at the time of the check.
The lemma itself is trivial, claiming only that at any given moment
either $A$ has not yet passed $B$,
or $A$ has passed $B$ but $B$ has not passed $A$,
or both agents have passed each other.

We also support defining a limited set of axioms 
(and/or previously proved lemmas) to use
when proving a specific lemma.
This reduces the proof search space
and therefore drastically improves the performance.
There are cases where specifying the proof dependencies
manually is easy;
in addition,
our lemma generation strategies include generation
of such dependency hints where appropriate.

\subsection{Additional proof-handling capabilities}

We use TPI (TPTP Process Instructions)
to specify operations on lemmas inside the proof.
To improve interactive usability, we allow a special declaration
that declares valid all the checked lemmas earlier in the proof.
This can be convenient to skip a part of the proof that has already
been verified earlier, or just to focus on a step in the middle 
of the proof before spending time on a possible unsuitable beginning.

In many cases, lemmas needed to achieve good performance
of the proof search are predictable.
Some of the techniques described in \cite{NASA} are broadly applicable,
especially proving all the components of each conjunction separately.
Another important 
source of lemmas is case-by-case analysis,
which requires choosing the cases but becomes 
a purely mechanical task
afterwards. 

For example, consider the following case.
The lemma under consideration claims that if it is impossible for two agents
to have passed each other, and two agents are distinct,
and reaching the critical section requires passing all the other agents,
then the two agents cannot both be in the critical section.
It is easier to prove the conclusion if we know whether some 
agent hasn't passed the other one (in which case we can say it has not
reached the critical section), or both agents have passed each other (in which
case we obtain a contradiction with impossibility of mutual passing).
So we prove exhaustiveness of a list of possible situations, 
and prove the lemma in each of them before proving it in the general case.

\begin{verbatim}
fof(safety_conditions_local_cases, checked_lemma,
  ![T,A,B]:( ~passed(T,A,B)
    | (passed(T,A,B) & ~passed(T,B,A))
    | (passed(T,A,B) & passed(T,B,A)))).

fof(safety_conditions_local_simplified,
  checked_lemma,
  ![T,A,B]:
  ((passed_exclusive_for(T,A,B) & A!=B
      & passed_in_critical_for(T,A)
      & passed_in_critical_for(T,B)) =>
      (active_state(T,A)!=criticalSection
      | active_state(T,B)!=criticalSection))).

tpi(ca_safety_conditions_local, add_cases,
  safety_conditions_local_cases =>
    safety_conditions_local_simplified).
\end{verbatim}

It will be checked that the case enumeration is exhaustive,
and then the main lemma will be checked with each of the cases
added as an additional assumptions, e.g.
\begin{verbatim}
fof(ca_safety_conditions_local_simplified_case_...,
  checked_lemma,
  ![T,A,B]:
  ((passed_exclusive_for(T,A,B) & A!=B
      & passed_in_critical_for(T,A)
      & passed_in_critical_for(T,B)
      & ~passed(T,A,B)) =>
      (active_state(T,A)!=criticalSection
      | active_state(T,B)!=criticalSection))).
\end{verbatim}

One more tool which sometimes unexpectedly turns out to be useful
is definition expansion. We have an instruction that expands 
specified definitions in a given formula. It turns out that there
are formulas that are simpler for existing provers if some definitions
are expanded.
Implementation of this functionality translates the definitions to expand
into code performing the expansion and uses the run-time 
code evaluation and compilation capabilities provided by Common Lisp
to run this code.

\subsection{Processing the prover output}

We have some tools for processing the output of automated theorem provers.
If a prover has produced a proof in the TPTP format,
it can be translated into
Graphviz (an automated graph layout tool) or
VUE (Visual Understanding Environment,
a GUI tool which includes functionality convenient 
for working with some types of graphs) format for visualization,
Visualization is supported both for the details of an individual lemma proof,
and for an overview of the global lemma dependence.

We also have a tool for detection of unused lemmas.
Unfortunately, sometimes removing unused lemmas from a proof makes
the task much harder for some of the provers.
The most likely reason for that is that even eventually unused axioms
affect the prioritization of possible directions of proof search.

\subsection{Parallel processing considerations}

Following the naive interpretation of upstream TPI semantics,
the operations on proofs are defined 
in imperative terms and operate on the entire proof.
This currently limits the opportunities for parallel execution
of the lemma-preprocessing code.
On the other hand, invoking external ATP systems provides an isolated
task to each prover instance,
and to aggregate the results we just need to check that every instance
has printed a line signalling successful proof.
We currently work with proofs in an interactive mode, 
observing the proof verification progress step by step.
To verify non-interactively a complete generated proof
our tooling allows to export all the prover tasks,
so that the prover invocations can be scheduled in any desired way
(possibly with multiple computers involved in processing).
We do not use this mode of operation yet.

\section{Conclusion}

We present a set of proof-manipulation tools 
that already implements quite a few useful operations
and will further grow in parallel with the research they support.

We currently use the tool set described in the present paper to explore
the performance implications of using different representations and using
different provers for distributed algorithms.
For example, we have encoded the inductive step proof for safety
of Dijkstra's mutual exclusion algorithm.
We plan to develop the capabilities further, supporting both
computer-assisted proof construction and providing an intermediate
representation for automated verification via proof generation.
We hope that the approach and some parts of our code might be of use to
others.
A mirror of the code is available at
\url{https://gitlab.common-lisp.net/mraskin/gen-fof-proof/}.

\begin{acks}
 The authors would like to thank the anonymous referees for
 their comments and suggestions about presentation.
\end{acks}


\begin{thebibliography}{88}
\bibitem{TPTP} Geoff Sutcliffe.
        The TPTP Problem Library and Associated Infrastructure.
                                    From CNF to TH0, TPTP v6.4.0
        Journal of Automated Reasoning, 2017, vol. 59, № 4, pp. 483--502.
\bibitem{TPI}
        Geoff Sutcliffe.
                The TPTP Process Instruction (TPI) Language.
        Retrieved on 18 March 2019.
        \url{http://tptp.cs.miami.edu/~tptp/TPTP/Proposals/TPILanguage.html}
\bibitem{NASA}     Ewen Denney, Bernd Fischer, Johann Schumann.
        Using Automated Theorem Provers to Certify Auto-generated Aerospace
          Software. International Joint Conference on Automated Reasoning 2004.
\bibitem{CASC} Geoff Sutcliffe. The CADE ATP System Competition - CASC.
        AI Magazine, 2016, vol. 37, № 2, pp. 99--101.
\bibitem{CASC-results} The CADE ATP System Competition --- 
        The World Championship for Automated Theorem Proving,
                homepage.
                Retrieved on 18 March 2019.
                \url{http://tptp.cs.miami.edu/~tptp/CASC/}
\bibitem{E}
        Stephan Schulz.
                System Description: E~1.8.
                Proc.\ of the 19th LPAR, Stellenbosch, 2013, pp. 735-743.
\bibitem{Esrap}
        Esrap project homepage.
        Retrieved on 28 January 2019.
                \url{http://nikodemus.github.com/esrap/}
\bibitem{packrat}
Bryan Ford, Packrat Parsing: a Practical Linear Time
    Algorithm with Backtracking. 2002.
        Retrieved on 28 January 2019.
                \url{http://pdos.csail.mit.edu/~baford/packrat/thesis/}
\bibitem{Dijkstra} E. W. Dijkstra.
  Solution of a problem in concurrent programming control.
    Commun. ACM 8, 9 (September 1965).
\bibitem{Coq} Coq Development Team. The Coq Proof Assistant Reference Manual,
                version 8.7. Retrieved on 27 January 2019.
                \url{http://coq.inria.fr}
\bibitem{CoqHammer}
        Ł. Czajka and C. Kaliszyk. Hammer for Coq: Automation for Dependent Type Theory.
                Journal of Automated Reasoning, 2018, vol.~61, issue 1--4, pp. 423--453.
        \url{http://cl-informatik.uibk.ac.at/cek/coqhammer/coqhammer.pdf}
\bibitem{Isabelle} Tobias Nipkow, Lawrence C Paulson, Markus Wenzel.
        Isabelle/HOL - A Proof Assistant for Higher-Order Logic
                Lecture Notes in Computer Science (2002).
\bibitem{Satallax} Chad E. Brown.
        Satallax: An Automatic Higher-Order Prover
\bibitem{graphviz}
        Graphviz project homepage.
        Retrieved on 18 March 2019.
                \url{https://graphviz.org/}
\bibitem{vue}
        Visual Understanding Environment project homepage.
        Retrieved on 18 March 2019.
                \url{https://vue.tufts.edu/}
\end{thebibliography}
\end{document}